\documentstyle[preprint,aps]{revtex}
\def\ie{{\it i.e.}}
\def\eg{{\it e.g.}}

\def\etal{{\it et al.}}

\def\beq{\begin{equation}}
\def\eeq{\end{equation}}
\def\epsb{\epsilon_b}
\def\tanb{\tan\beta}

\def\bsg{b\to s\gamma}
\def\Zbb{Z\rightarrow b\ov b}
\def\sub#1{_{\lower.25ex\hbox{$\scriptstyle#1$}}}
\def\sul#1{_{\kern-.1em#1}}
\def\sll#1{_{\kern-.2em#1}}
\def\sbl#1{_{\kern-.1em\lower.25ex\hbox{$\scriptstyle#1$}}}
\def\ssb#1{_{\lower.25ex\hbox{$\scriptscriptstyle#1$}}}
\def\sbb#1{_{\lower.4ex\hbox{$\scriptstyle#1$}}}

\def\GeV{\,{\rm GeV}}

\def\JL{J. L. Lopez}
\def\DVN{D. V. Nanopoulos}

\def\r#1{$\bf#1$}
\def\rb#1{$\bf\overline{#1}$}
\def\to{\rightarrow}
\def\ov{\overline}

\def\GeV{\,{\rm GeV}}
\def\TeV{\,{\rm TeV}}
\def\mh{\ifmmode m\sbl H \else $m\sbl H$\fi}
\def\mch{\ifmmode m_{H^\pm} \else $m_{H^\pm}$\fi}
\def\mt{\ifmmode m_t\else $m_t$\fi}
\def\mc{\ifmmode m_c\else $m_c$\fi}
\def\mz{\ifmmode M_Z\else $M_Z$\fi}
\def\mw{\ifmmode M_W\else $M_W$\fi}
\def\mws{\ifmmode M_W^2 \else $M_W^2$\fi}
\def\mhs{\ifmmode m_H^2 \else $m_H^2$\fi}
\def\mzs{\ifmmode M_Z^2 \else $M_Z^2$\fi}
\def\mts{\ifmmode m_t^2 \else $m_t^2$\fi}
\def\mcs{\ifmmode m_c^2 \else $m_c^2$\fi}
\def\mchs{\ifmmode m_{H^\pm}^2 \else $m_{H^\pm}^2$\fi}
\def\ztwo{\ifmmode Z_2\else $Z_2$\fi}
\def\zone{\ifmmode Z_1\else $Z_1$\fi}
\def\mtwo{\ifmmode M_2\else $M_2$\fi}
\def\mone{\ifmmode M_1\else $M_1$\fi}
\def\tb{\ifmmode \tan\beta \else $\tan\beta$\fi}
\def\xw{\ifmmode x\sub w\else $x\sub w$\fi}
\def\ch{\ifmmode H^\pm \else $H^\pm$\fi}
\def\lum{\ifmmode {\cal L}\else ${\cal L}$\fi}
\def\inpb{\ifmmode {\rm pb}^{-1}\else ${\rm pb}^{-1}$\fi}
\def\infb{\ifmmode {\rm fb}^{-1}\else ${\rm fb}^{-1}$\fi}
\def\epem{\ifmmode e^+e^-\else $e^+e^-$\fi}
\def\ppb{\ifmmode \bar pp\else $\bar pp$\fi}

\newskip\zatskip \zatskip=0pt plus0pt minus0pt

\def\matth{\mathsurround=0pt}
\def\lsim{\mathrel{\mathpalette\atversim<}}
\def\gsim{\mathrel{\mathpalette\atversim>}}
\def\atversim#1#2{\lower0.7ex\vbox{\baselineskip\zatskip\lineskip\zatskip
  \lineskiplimit 0pt\ialign{$\matth#1\hfil##\hfil$\crcr#2\crcr\sim\crcr}}}

\begin{document}
\preprint{\begin{tabular}{c}
\hbox to\textwidth{August 1994 \hfill SNUTP 94-66}\\[-10pt]
\hbox to\textwidth{hep-ph/9408218 \hfill}\\[36pt]
\end{tabular}}
\draft
\title{$\epsb$ Constraints on the Minimal $SU(5)$\\
 and $SU(5)\times U(1)$ Supergravity Models}
\author{Jihn~E.~Kim$^{(a),(b)}$ and Gye~T.~Park$^{(a)}$}
\address {\it $^{(a)}$Center for Theoretical Physics and
$^{(b)}$Department of Physics\\
%[-8pt]
Seoul National University, Seoul, 151-742, Korea}
%[-8pt]

\date{\today}
\maketitle
\begin{abstract}

We have performed a systematic analysis
to compute the one-loop electroweak corrections to the $\Zbb$ vertex
in terms of $\epsilon_b$ and $R_b$ in the context of the minimal $SU(5)$ and
no-scale $SU(5)\times U(1)$ supergravity models.
With the measured top mass, $m_t=174\pm 10^{+13}_{-12}$ $\GeV$
, recently announced by CDF, we use the latest LEP data on $\epsilon_b$ and
$R_b$ ($\equiv{\Gamma(Z\rightarrow b\ov b)\over{\Gamma(Z\rightarrow
hadrons)}}$) in order to constrain further the two models.
We find that the present LEP data on $\epsilon_b$ and $R_b$ constrain the two
models rather severely. Especially, the low-$\tanb$ region is constrained more
severely.
$\tanb\lsim 2.5$ $(4.0)$ is excluded by $\epsb$ at $90\%$ C.~L. for $m_t\gsim
170$ $(180)$ $\GeV$ in the minimal $SU(5)$ (no-scale $SU(5)\times U(1)$)
supergravity. Even more stringent constraint comes from $R_b$.
It excludes $\tanb\lsim 4.0$ at $90\%$ C.~L. for $m_t\gsim 160$ $(170)$ $\GeV$
in the minimal $SU(5)$ (no-scale $SU(5)\times U(1)$) supergravity.
We also find that the sign on $\mu$ in the two models can be determined from
$\epsilon_b$ and $R_b$ at $90\%$ C.~L.

\end{abstract}

\narrowtext

\section{Introduction}
With the increasing accuracy of the LEP measurements, it has become more
important than ever performing the precision test of the standard model (SM)
and its extensions.
A standard model fit to the latest LEP data yields the top mass,
$m_t=178\pm 11^{+18}_{-19}$ $\GeV$  \cite{Schaile}.
With this large top quark mass, the $\Zbb$ vertex contribution, which is
proportional to $m_t^2$, becomes more significant, and can provide
a powerful tool to constrain $m_t$ experimetally. This is still very useful
because
the measured top mass from CDF  \cite{CDF-top}, $m_t=174\pm 10^{+13}_{-12}$
$\GeV$, has a large error bars and D0 gives just the lower bound on $m_t$,
$m_t\gsim 131$ $\GeV$ \cite{D0}.
With the improved measurement for the $Z$ partial width to $b\ov b$, primarily
due to the use of new life time-based techniques, one may be able to put more
precise bound on $m_t$.
The experimental value for $\Gamma(Z\rightarrow b\ov b)$ has increased over the
year, resulting in larger experimental value for $R_b$
($\equiv{\Gamma(Z\rightarrow b\ov b)\over{\Gamma(Z\rightarrow hadrons)}}$), and
therefore rather small upper bound on $m_t$  is favored in the SM
\cite{Schaile}. One could certainly
interpret this as a possible manifestation of new physics beyond the SM, where
at one loop the negative standard top quark contributions are cancelled to a
certain extent by the contributions from the new particles, thereby allowing
considerably larger $m_t$ than in the SM.
In fact, the minimal supersymmetric standard model (MSSM) realizes this
possibility.

Another very interesting observable which encodes the one loop corrections
to the $\Zbb$ vertex is $\epsb$ first introduced in Ref.~\cite{ABC}.
In supergravity(SUGRA) models, radiative electroweak symmetry breaking
mechanism \cite{EWx} can be described by at most 5 parameters: the top-quark
mass ($m_t$), the ratio of Higgs
vacuum expectation values ($\tanb$), and three universal
soft-supersymmetry-breaking parameters ($m_{1/2},m_0,A$) \footnote{See,
however, Ref.~\cite{non-univ} for
non-universal soft-supersymmetry breaking parameters}.

In this work we explore the minimal $SU(5)$ SUGRA  \cite{su5sugra} and the
no-scale $SU(5)\times U(1)$
SUGRA  \cite{flippedsugra} in terms of $\epsb$ parameter which encodes the
one-loop corrections to the $\Zbb$ vertex. Moreover, we attempt to see how well
these models can fit in
rather uncomfortably high 1993 LEP value for $R_b$ ($\equiv{\Gamma(Z\rightarrow
b\ov b)\over{\Gamma(Z\rightarrow hadrons)}}$).

\section{The minimal $SU(5)$ and $SU(5)\times U(1)$ SUGRA models}
The minimal $SU(5)$ and $SU(5)\times U(1)$ SUGRA models both contain, at low
energy, the
SM gauge symmetry and the particle content of the MSSM. A few crucial
differences between the two models are: \\
(i) The unification groups are different, $SU(5)$ versus $SU(5)\times U(1)$.\\
(ii) The gauge coupling unification occurs at
$\sim 10^{16}$ $\GeV$ in the minimal $SU(5)$ model whereas in $SU(5)\times
U(1)$ model it occurs at the string scale $\sim 10^{18}$ $\GeV$
\cite{LNZstring}.
In $SU(5)\times U(1)$ SUGRA, the gauge unification is delayed because of the
effects of an additional pair of \r{10},\rb{10} vector-like representations
with intermediate-scale masses. The different heavy field content at the
unification scale leads to different constraints from proton decay. \\
(iii) In the minimal $SU(5)$
SUGRA, proton decay is highly constraining whereas it is not in $SU(5)\times
U(1)$ SUGRA.

The procedure to restrict 5-dimensional parameter spaces is as follows
\cite{aspects}.
First, upon sampling a specific choice of ($m_{1/2},m_0,A$) at the unification
scale and ($m_t,\tanb$) at the electroweak scale, the renormalization group
equations (RGE) are run from the unification scale to the electroweak scale,
where the radiative electroweak breaking condition is imposed by minimizing the
effective 1-loop Higgs potential, which determines the Higgs mixing term $\mu$
up to its sign.
We also impose consistency constraints such as perturbative unification and the
naturalness bound of $m_{\tilde g}\lsim 1\TeV$.
Finally, all the known experimental bounds on the sparticle masses are imposed
\footnote{We use the following experimental lower bounds on the sparticle
masses in GeV in the order of gluino, squarks, lighter stop, sleptons, and
lighter chargino: $m_{\tilde g}\gsim 150$, $m_{\tilde q}\gsim 100$,
$m_{{\tilde{t}}_1}\gsim 45$, $m_{\tilde l}\gsim 43$,
$m_{\chi^\pm_1}\gsim 45$.}.
This prodedure yields the restricted parameter spaces for the two models.

Further reduction in the number of input parameters in $SU(5)\times U(1)$ SUGRA
is made possible because in specific string-inpired scenarions for
($m_{1/2},m_0,A$) at the unification scale these three parameters are computed
in terms of just one of them \cite{IL}. One obtains $m_0=A=0$ in the {\em
no-scale} scenario and
$m_0=\frac{1}{\sqrt{3}}m_{1/2}$, $A=-m_{1/2}$ in the {\em dilaton} scenario
\footnote{Note, however, that one loop correction changes this relation
significantly \cite{cosCKN}.}.

The low energy predictions for the sparticle mass spectra are quite different
in the two SUGRA models mainly due to the different pattern of supersymmetry
radiative breaking.
In the minimal $SU(5)$ SUGRA, all the squarks except the lighter stop and
all the Higgs except the lighter neutral Higgs are quite heavy ( $\gsim$ a few
hundred GeV ) whereas they can be quite light in the $SU(5)\times U(1)$ SUGRA.
This difference leads to strikingly different phenomenology in the two models,
for example in the flavor changing radiative decay $\bsg$  \cite{bsgLNP}.

\section{One-loop electroweak radiative corrections and the new $\epsilon$
parameters}
There are several schemes to parametrize the electroweak vacuum
polarization corrections  \cite{Kennedy,PT,efflagr,AB}. It can be shown, by
expanding the vacuum polarization tensors to order $q^2$, that one obtains
three independent physical parameters. Alternatively, one can show that upon
symmetry breaking three additional terms appear in the effective lagrangian
 \cite{efflagr}. In the $(S,T,U)$ scheme  \cite{PT}, the deviations of the
model
predictions from the SM predictions (with fixed SM values for $m_t,m_{H_{SM}}$)
are considered as the effects from ``new physics". This scheme is only valid to
the lowest order in $q^2$, and is therefore not applicable to a theory with
light new particles comparable to $M_Z$. In the $\epsilon$-scheme
\cite{ABC,ABJ}, on
the other hand, the model predictions are absolute and also valid up to higher
orders in $q^2$, and therefore this scheme is more applicable to the
electroweak
precision tests of the MSSM  \cite{BFC} and a class of supergravity models
 \cite{ewcorr}.

There are two different $\epsilon$-schemes. The original scheme \cite{ABJ} was
considered in one of author's previous analyses  \cite{ewcorr,bsg-eps}, where
$\epsilon_{1,2,3}$ are defined from a basic set of observables $\Gamma_{l},
A^{l}_{FB}$ and $M_W/M_Z$.
Due to the large $m_t$-dependent vertex corrections to $\Gamma_b$, the
$\epsilon_{1,2,3}$ parameters   and $\Gamma_b$ can be correlated only for a
fixed value of $m_t$. Therefore, $\Gamma_{tot}$, $\Gamma_{hadron}$ and
$\Gamma_b$ were not included  in Ref.~\cite{ABJ}. However, in the new
$\epsilon$-scheme, introduced recently in Ref.~\cite{ABC}, the above
difficulties are overcome by introducing a new parameter $\epsilon_b$ to encode
the $\Zbb$ vertex corrections. The four $\epsilon$'s are now defined from an
enlarged set of $\Gamma_{l}$, $\Gamma_{b}$, $A^{l}_{FB}$ and $M_W/M_Z$ without
even specifying $m_t$.
This new scheme was adopted in a previous analysis by one of us (G.P.) in the
context of the $SU(5)\times U(1)$ SUGRA models \cite{LNPZepsb}.
In this work we use this new $\epsilon$-scheme.
As is well known, the SM contribution to $\epsilon_1$ depends quadratically
on $m_t$ but only logarithmically on the SM Higgs boson mass ($m_H$). Therefore
upper bounds on $m_t$  have a non-negligible $m_H$
dependence: up to $20$ $\GeV$ stronger when going from a heavy ($\approx1\TeV$)
to a light ($\approx100$ $\GeV$) Higgs boson. It is also known in the MSSM that
the largest supersymmetric contributions to $\epsilon_1$ are expected to
arise from the $\tilde t$-$\tilde b$ sector, and in the limiting case of a very
light stop, the contribution is comparable to that of the $t$-$b$ sector. The
remaining squark, slepton, chargino, neutralino, and Higgs sectors all
typically contribute considerably less. For increasing sparticle masses, the
heavy sector of the theory decouples, and only SM effects  with a {\it light}
Higgs boson survive. However, for a
light chargino ($m_{\chi^\pm_1}\to{1\over2}M_Z$), a $Z$-wavefunction
renormalization threshold effect coming from Z-vacuum polarization diagram with
the lighter chargino in the loop
 can introduce a substantial $q^2$-dependence
in the calculation  \cite{BFC}.
This results in a weaker upper bound on $m_t$ than in the SM.
The complete vacuum polarization contributions from the Higgs sector, the
supersymmetric chargino-neutralino and sfermion sectors, and also the
corresponding contributions in the SM have been included in our calculations
 \cite{ewcorr}. However, the supersymmetric contributions to the non-oblique
corrections except in $\epsb$ have been neglected.

Following Ref.~\cite{ABC}, $\epsb$ is defined from $\Gamma_b$, the inclusive
partial width for $\Zbb$, as
\begin{equation}
\epsb={g^b_A\over{g^l_A}}-1
\end{equation}
where $g^b_A$ $(g^l_A)$ is the axial-vector coupling of $Z$ to $b$ $(l)$.
In the SM, the diagrams for $\epsb$  involve top quarks and
$W^\pm$ bosons  \cite{RbSM}, and the contribution to $\epsb$ depends
quadratically on $m_t$ ($\epsb=-G_F m_t^2/4\sqrt {2}\pi^2 + \cdots$).
In supersymmetric models there are additional diagrams
involving Higgs bosons and supersymmetric particles. The charged Higgs
contributions have been calculated in Refs.~ \cite{Denner,epsb2HD} in
the context of a non-supersymmetric two Higgs doublet model, and the
contributions involving supersymmetric particles in Refs.~ \cite{BF,Rb2HD}.
The main features of the additional supersymmetric contributions are: (i) a
negative contribution
from charged Higgs--top exchange which grows as $m^2_t/\tan^{2}\beta$ for
$\tan\beta\ll{m_t\over{m_b}}$; (ii) a positive contribution from chargino-stop
exchange which in this case grows as $m^2_t/\sin^{2}\beta$; and (iii) a
contribution from neutralino(neutral Higgs)--bottom exchange which grows as
$m^2_b\tan^{2}\beta$ and is negligible except for large values of $\tan\beta$
(\ie, $\tan\beta\gsim{m_t\over{m_b}}$) (the contribution (iii) has been
neglected in our analysis).

\section{Results and discussion}
In Figure 1 we present our numerical results for $\epsb$ in the two SUGRA
models. $\alpha_S(M_Z)=0.118$ and $m_b=4.8$ $\GeV$ are used throughout the
numerical calculations.
We use the experimental value for $\epsb$,
$\epsb^{exp}=(0.9\pm 4.2)\times 10^{-3}$, determined from the latest
$\epsilon$- analysis using the LEP and SLC data in Ref.~\cite{Barbieri94}.
The discontinuity in the chargino mass in the minimal model in the figure
is simply due to the use of
large steps in sampling the value of $m_{1/2}$.
The values of $m_t$ are chosen in such a way that the approximate
$\epsb$-deduced $m_t$ bounds are readily obtained from the figure. Only one
value of
$m_t$ is displayed in the $SU(5)\times U(1)$ SUGRA because considerable portion
of the model predictions are overlapped for two different values of $m_t$
due to the steep rise in $\epsb$ for a light chargino.
The reason why the rise in $\epsb$ in the $SU(5)\times U(1)$ SUGRA
is much steeper than in the minimal $SU(5)$ SUGRA is that
the stop mass scales with the chargino mass in the no-scale model
whereas it does not in the minimal model.
Therefore, the light chargino effect in $\epsb$ is optimized better
in the no-scale $SU(5)\times U(1)$ SUGRA.
%the heavier stop in the minimal $SU(5)$ SUGRA can only be very heavy.
This difference leads to different $\epsb$-deduced $m_t$ bounds
in the two models.
The approximate bounds at $90\% $ C.~L. are $m_t\lsim 175$ $(185)$ $\GeV$ for
the minimal $SU(5)$ (no-scale $SU(5)\times U(1)$) SUGRA.
In the no-scale model, one can also determine the sign on $\mu$ to be positive
for $m_t\gsim 180$ $\GeV$.
The lowest value of $\epsb$ for a fixed $m_t$ represents the lowest $\tanb$ for
not too large $\tanb$ \footnote{For large $\tan\beta(\gsim {m_t\over{m_b}})$,
the charged Higgs diagram gets a significant contribution proportional to
$-m^2_b\tan^{2}\beta$ coming from the charged Higgs coupling to $b_R$, thereby
driving $\epsb$ even below the value corresponding to the lowest $\tanb$.}.
It is $\tanb=1.5$ $(4.0)$
for the minimal $SU(5)$ (no-scale $SU(5)\times U(1)$) SUGRA in Figure 1.
{}From this, we obtain low $\tanb-m_t$ correlated bounds at $90\%$ C.~L., which
are
for $\tanb\lsim 2.5$ $(4.0)$, $m_t\lsim 170 (180)$ $\GeV$ in the minimal
$SU(5)$ (no-scale $SU(5)\times U(1)$) SUGRA.
Although the $m_t$ values from CDF have rather large error bars at present, one
can imagine an interesting situation in the near future where the $m_t$ values
from CDF turns out to fall between the above $\epsb$-deduced $m_t$ bounds,
disfavoring only one model.

The experimental value for $R_b$($\equiv{\Gamma(Z\rightarrow b\ov
b)\over{\Gamma(Z\rightarrow hadrons)}}$) from the 1993 LEP data are reported
very recently
to be rather  high, $0.2192\pm 0.0018$, in comparison with the SM predictions
\cite{Schaile}.
In an attempt to see how much the situation can improve in SUGRA models,
we now calculate $R_b$ in the two SUGRA models \footnote{We use the expression
for $R_b$ in terms of $\epsilon$'s given in Ref.~\cite{ABC}}. In Figure 2 we
show the model predictions for $R_b$ in the two models.
As seen in the figure, the $R_b$ constraint is much stronger than the $\epsb$
constraint. The $R_b$-deduced $m_t$ bounds at $90\%$ C.~L. are
$m_t\lsim 165$ $(175)$ $\GeV$ for the minimal $SU(5)$ (no-scale $SU(5)\times
U(1)$) SUGRA.
{}From the figure, one can also put bounds on the chargino mass, which
are  $m_{\chi^\pm_1}\lsim 85$ $(70)$ $\GeV$ for $m_t\gsim 160$ $(170)$ $\GeV$
for the minimal $SU(5)$ (no-scale $SU(5)\times U(1)$) SUGRA.
Similarly, one can also obtain bounds on the lighter stop mass given by
$m_{{\tilde{t}}_1}\lsim 500$ $(190)$ $\GeV$ for $m_t\gsim 160$ $(170)$ $\GeV$
for the minimal $SU(5)$ (no-scale $SU(5)\times U(1)$) SUGRA.
Therefore, in the minimal $SU(5)$ (no-scale $SU(5)\times U(1)$) SUGRA, if the
top turns out to be heavier than $160$ $(170)$ $\GeV$,
then only the lighter chargino may be detected at LEPII.
The $\tanb$-dependence is very pronounced in the no-scale model for $\mu >0$.
The low values of $\tanb$ are as indicated in the figure.
For $\tanb=2$, the dotted curve becomes nearly flat as the chargino mass
becomes large. This is because the
charged Higgs contribution nearly cancels the chargino contribution  \cite{BF},
making $R_b$ get saturated much faster to the SM value.
As in the $\epsb$ constraint above, the low $\tanb-m_t$ correlated bounds at
$90\%$ C.~L.
are obtained as follows: for $\tanb\lsim 4.0$, $m_t\lsim 160$ $(170)$ $\GeV$ in
the minimal $SU(5)$ (no-scale $SU(5)\times U(1)$) SUGRA.
In the no-scale model, $\tanb\lsim 2$ is excluded even at $95\%$ C.~L.
for $m_t\gsim 170$ $\GeV$. From $R_b$, one can also determine $\mu$
to be positive in both models.
It is very interesting for one to see that the low-$\tanb$ region is severely
constrained by both constraints above.
We would like to stress here the fact that our calculations are fairly accurate
in the low-$\tanb$ region because the diagrams neglected in the calculations
can be safely neglected there.
The major features of the constraints from $\epsb$ and $R_b$ for the two SUGRA
models are summarized in the Table 1.

\section{Conclusions}

We have computed the one-loop electroweak corrections to the $\Zbb$ vertex
in terms of $\epsilon_b$ and $R_b$ in the context of the minimal $SU(5)$ and
no-scale $SU(5)\times U(1)$ supergravity models. We use the latest LEP data on
$\epsilon_b$ and $R_b$
in order to constrain further the two models.
We find that the present LEP data on $\epsilon_b$ and $R_b$ constrain the two
models rather severely. Especially, the low-$\tanb$ region is constrained more
severely.
$\tanb\lsim 2.5$ $(4.0)$ is excluded by $\epsb$ at $90\%$ C.~L. for $m_t\gsim
170$ $(180)$ $\GeV$ in the minimal $SU(5)$ (no-scale $SU(5)\times U(1)$) SUGRA.
Even more stringent constraint comes from $R_b$.
It excludes $\tanb\lsim 4.0$ at $90\%$ C.~L. for $m_t\gsim 160$ $(170)$ $\GeV$
in the minimal $SU(5)$ (no-scale $SU(5)\times U(1)$) SUGRA.
We also find that the sign on $\mu$ in the two models can be determined from
$\epsilon_b$ and $R_b$ at $90\%$ C.~L.
This can be of special interest in the minimal $SU(5)$ because
the low-$\tanb$ region is phenomenologically favored by the measured
ratio $m_b/{m_\tau}$. We also find that the sign on $\mu$ in the two models can
be determined from $\epsilon_b$ and $R_b$ at $90\%$ C.~L.

With improved measurement on the top mass by CDF in the near future,
there may be an amusing possibility that one could favor one model over
the other from the $\Zbb$ constraints.
And also, in the no-scale $SU(5)\times U(1)$ SUGRA, if the top turns out to be
heavier than $170$ $\GeV$,
then only the lighter chargino lighter than $80$ $\GeV$ may be detected at
LEPII.

\acknowledgments
This work has been supported in part by the Korea Science and Engineering
Foundation through Center for Theoretical Physics, Seoul National University.
JEK is also supported by the Ministry of Education through the Basic Science
Research Institute, Contract No. BSRI-94-2418.

\def\NPB#1#2#3{Nucl. Phys. B {\bf#1} (19#2) #3}
\def\PLB#1#2#3{Phys. Lett. B {\bf#1} (19#2) #3}
\def\PLIBID#1#2#3{B {\bf#1} (19#2) #3}
\def\PRD#1#2#3{Phys. Rev. D {\bf#1} (19#2) #3}
\def\PRL#1#2#3{Phys. Rev. Lett. {\bf#1} (19#2) #3}
\def\PRT#1#2#3{Phys. Rep. {\bf#1} (19#2) #3}
\def\MODA#1#2#3{Mod. Phys. Lett. A {\bf#1} (19#2) #3}
\def\IJMP#1#2#3{Int. J. Mod. Phys. A {\bf#1} (19#2) #3}
\def\TAMU#1{Texas A \& M University preprint CTP-TAMU-#1}
\def\ARAA#1#2#3{Ann. Rev. Astron. Astrophys. {\bf#1} (19#2) #3}
\def\ARNP#1#2#3{Ann. Rev. Nucl. Part. Sci. {\bf#1} (19#2) #3}

\begin{figure}
\caption{The predictions for $\epsilon_b$
 versus the lighter chargino mass in the minimal $SU(5)$ SUGRA for
 $m_t=160, 175$ $\GeV$ (top row) and in the no-scale $SU(5)\times
U(1)$ SUGRA for $m_t=180$ $\GeV$ (bottom row).
The values of $m_t$ are as indicated.
The points above the horizontal solid line are allowed at 90\% C.L.}
\label{epsb1}
\end{figure}
\begin{figure}
\caption{The predictions for $R_b$
 versus the lighter chargino mass in the minimal $SU(5)$ SUGRA for
 $m_t=160$ $\GeV$ (top row) and in the no-scale $SU(5)\times
U(1)$ SUGRA for $m_t=170$ $\GeV$ (bottom row).
The values of $\tanb$ are as indicated near the dotted curves (bottom row).
The points above the horizontal solid lines are allowed at 90 or 95\% C.L.}
\label{epsb2}
\end{figure}
\begin{figure}
\end{figure}

% \begin{table}
% \caption{}
% \label{}
% \end{tabular}
% \end{table}
%\newpage
\begin{table}
\hrule
\caption{The major features of the constraints from $\epsb$ and $R_b$ for the
two SUGRA models considered.}
\label{Table1}
\begin{center}
\begin{tabular}{|c|c|c|}\hline
&Minimal $SU(5)$&no-scale $SU(5)\times U(1)$\\ \hline
$\epsb$ (90\% C.L.)&$m_t\lsim 175$ $\GeV$ for any $\tanb$&$m_t\lsim 185$ $\GeV$
for any $\tanb$\\
&$m_t\lsim 170$ $\GeV$ for $\tanb\lsim 2.5$&$m_t\lsim 180$ $\GeV$ for
$\tanb\lsim 4$\\
$R_b$ (90\% C.L.)&$m_t\lsim 165$ $\GeV$ for any $\tanb$&$m_t\lsim 175$ $\GeV$
for any $\tanb$\\
&$m_t\lsim 160$ $\GeV$ for $\tanb\lsim 4$&$m_t\lsim 170$ $\GeV$ for $\tanb\lsim
4$\\
&For $m_t\gsim 160$ $\GeV$,  &For $m_t\gsim 170$ $\GeV$, \\
& $m_{\chi^\pm_1}\lsim 85$ $\GeV$ and $m_{{\tilde{t}}_1}\lsim 500$ $\GeV$ &
$m_{\chi^\pm_1}\lsim 70$ $\GeV$ and $m_{{\tilde{t}}_1}\lsim 190$ $\GeV$\\
\hline
\end{tabular}
\end{center}
\hrule
\end{table}


\begin{references}
\bibitem{Schaile} D. Schaile, talk given at 27th International Conference
on High Energy Physics, Glasgow, July 1994.
\bibitem{CDF-top} CDF Collaboration, \PRL{73}{94}{225}.
\bibitem{D0} D0 Collaboration, \PRL{72}{94}{2138}.
\bibitem{ABC}G. Altarelli, R. Barbieri, and F. Caravaglios, \NPB{405}{93}{3}.
\bibitem{EWx}L. Ib\'a\~nez and G. Ross, \PLB{110}{82}{215}; K. Inoue, \etal,
Prog. Theor. Phys. 68 (1982) 927; L. Ib\'a\~nez, \NPB{218}{83}{514} and
\PLB{118}{82}{73}; H. P. Nilles, \NPB{217}{83}{366}; J. Ellis, \DVN, and
K. Tamvakis, \PLB{121}{83}{123}; J. Ellis, J. Hagelin, \DVN, and K. Tamvakis,
\PLB{125}{83}{275}; L. Alvarez-Gaum\'e, J. Polchinski, and M. Wise,
\NPB{221}{83}{495}; L. Iba\~n\'ez and C. L\'opez, \PLB{126}{83}{54} and
\NPB{233}{84}{545}; C. Kounnas, A. Lahanas, \DVN, and M. Quir\'os,
\PLB{132}{83}{95} and C. Kounnas, A. Lahanas, \DVN, and M. Quir\'os,
\NPB{236}{84}{438}.
\bibitem{non-univ} D. Matalliotakis and H.~P.~Nilles, TUM-HEP-201/94.
\bibitem{su5sugra} For reviews see R. Arnowitt and P. Nath, {\it Applied N=1
Supergravity} (World Scientific, Singapore 1983);
H. P. Nilles, \PRT{110}{84}{1}.
\bibitem{flippedsugra} For a recent review see \JL, \DVN, and A. Zichichi,
CERN-TH.6926/93 (unpublished).
\bibitem{LNZstring} \JL, \DVN, and A. Zichichi, \PRD{49}{94}{343} and
references therein.
\bibitem{aspects} S. Kelley, \JL, \DVN, H. Pois, and K. Yuan,
\NPB{398}{93}{3}.
\bibitem{IL}See \eg, L. Ib\'a\~nez and D. L\"ust, \NPB{382}{92}{305};
V. Kaplunovsky and J. Louis, \PLB{306}{93}{269}; A. Brignole, L. Ib\'a\~nez,
and C. Mu\~noz, FTUAM-26/93.
\bibitem{cosCKN} K.~Choi, J.~E.~Kim and H.~P.~Nilles, SNUTP-94-19 (1994).
\bibitem{bsgLNP}\JL, \DVN, and G.~T. Park, \PRD{48}{93}{R974}.
\bibitem{Kennedy} D. Kennedy and B. Lynn, \NPB{322}{89}{1};
D. Kennedy, B. Lynn, C. Im, and R. Stuart, \NPB{321}{89}{83}.
\bibitem{PT} M. Peskin and T. Takeuchi, \PRL{65}{90}{964};
W. Marciano and J. Rosner, \PRL{65}{90}{2963};
D. Kennedy and P. Langacker, \PRL{65}{90}{2967}.
\bibitem{efflagr} B. Holdom and J. Terning, \PLB{247}{90}{88};
M. Golden and L. Randall, \NPB{361}{91}{3};
A. Dobado, D. Espriu, and M. Herrero, \PLB{255}{91}{405}.
\bibitem{AB}G. Altarelli and R. Barbieri, \PLB{253}{90}{161}
\bibitem{ABJ}G. Altarelli, R. Barbieri, and S. Jadach, \NPB{369}{92}{3}.
\bibitem{BFC}R. Barbieri, M. Frigeni, and F. Caravaglios, \PLB{279}{92}{169}.
\bibitem{ewcorr}\JL, \DVN, G.~T. Park, H. Pois, and K. Yuan,
\PRD{48}{93}{3297}.
\bibitem{bsg-eps} \JL, \DVN, G.~T. Park, and A. Zichichi, \PRD{49}{94}{355}.
\bibitem{LNPZepsb} \JL, \DVN, G.~T. Park, and A. Zichichi, \PRD{49}{94}{4835}.
\bibitem{RbSM} J. Bernabeu, A. Pich, and A. Santamaria, \PLB{200}{88}{569};
 W. Beenaker and W. Hollik, Z. Phys. C40, 141(1988); A. Akhundov, D. Bardin,
and T. Riemann, \NPB{276}{86}{1}; F. Boudjema, A. Djouadi, and C. Verzegnassi,
  \PLB{238}{90}{423}.
\bibitem{Denner} A. Denner, R. Guth, W. Hollik, and J. K\"uhn,
Z. Phys. C51, 695(1991). The neutral Higgs contributions to $\Zbb$ were also
calculated here.
%\bibitem{Rbbsg2HD} G.~T.~Park, \PRD{49}{94}{}(July issue).
\bibitem{epsb2HD} G.~T.~Park, \MODA{9}{94}{321}.
\bibitem{BF} M. Boulware and D. Finnell, \PRD{44}{91}{2054}.
\bibitem{Rb2HD} A. Djouadi, G. Girardi, C. Verzegnassi, W. Hollik, and F.
Renard, \NPB{349}{91}{48}.
\bibitem{Barbieri94} R.~Barbieri, talk given at Rencontres Physique de la Valle
d'Aosta, La Thuile, IFUP-TH 28/94 (1994).

\end{references}
\end{document}